\documentclass[preprint,12pt,authoryear]{elsarticle}
\usepackage{booktabs}
\usepackage{hyperref}
\usepackage{threeparttable}

\usepackage{amssymb}
\usepackage{amsmath}
\usepackage{multirow}
\usepackage{xcolor}
\usepackage{url}

\journal{}

\begin{document}
\begin{frontmatter}
\title{A Detailed Study on LLM Biases Concerning Corporate Social Responsibility and Green Supply Chains}
\author[a]{Greta Ontrup\fnref{b}\corref{cor1}}
\ead{greta.ontrup@uni-due.de}
\author[a]{Annika Bush\fnref{c}}
\author[a]{Markus Pauly\fnref{d}}
\author[a]{Meltem Aksoy\fnref{c}}
\cortext[cor1]{Corresponding author}
\affiliation[a] {organization=Research Center Trustworthy Data Science and Security, University Alliance Ruhr}
\affiliation[b] {organization=Department of Computer Science, University of Duisburg-Essen}
\affiliation[c] {organization=Department of Computer Science, Technical University Dortmund}
\affiliation[d] {organization=Chair of Mathematical Statistics and Applications in Industry, Technical University Dortmund}

\begin{abstract}


Organizations increasingly use Large Language Models (LLMs) to improve supply chain processes and reduce environmental impacts. However, LLMs have been shown to reproduce biases regarding the prioritization of sustainable business strategies. Thus, it is important to identify underlying training data biases that LLMs pertain regarding the importance and role of sustainable business and supply chain practices.
This study investigates how different LLMs respond to validated surveys about the role of ethics and responsibility for businesses, and the importance of sustainable practices and relations with suppliers and customers. Using standardized questionnaires, we systematically analyze responses generated by state-of-the-art LLMs to identify variations. We further evaluate whether differences are augmented by four organizational culture types, thereby evaluating the practical relevance of identified biases. The findings reveal significant systematic differences between models and demonstrate that organizational culture prompts substantially modify LLM responses. The study holds important implications for LLM-assisted decision-making in sustainability contexts.
\end{abstract}


\end{frontmatter}

\section{Introduction}

Large language models (LLMs) have become increasingly influential in organizational decision-making. A specific application field is green supply chain management. Here, LLMs are adopted to enhance the sustainability of operations, assist in the preparation of sustainability reports, or support supplier evaluations \citep{zhou2025}. Other examples cover strategic planning, stakeholder communication, informing policy development, and reporting. 
The widespread adoption of LLMs within organizational settings prompts critical questions regarding how these models interpret and convey principles of sustainability.
LLMs are developed using extensive text-based datasets, which inherently mirror the societal values, cultural conventions, and biases contained within their training materials \citep{gallegos2024biasfairnesslargelanguage, aksoy2024}. In addition to training data, other factors such as the underlying model architecture \citep{weber2024gpt4politicallybiasedgpt35}, fine-tuning procedures \citep{Ferrara_2023}, instruction-tuning strategies \citep{resnik2025llmbias}, and the specific prompts used to elicit responses \citep{kamruzzaman2024prompting, brucks2025prompt} also shape model behavior. When deployed for green supply chain management, these embedded aspects can influence how organizations approach environmental management and social responsibility.

A recent study found that LLMs exhibit biases related to sustainable development \citep{bush2025emnlp-arxiv}. For instance, GPT-4o \citep{openai2024gpt4o} responses mirrored skepticism regarding the compatibility between sustainability and AI, while the output by LLaMA reflected techno-optimism \citep{bush2025emnlp-arxiv}. This demonstrates that an organization's choice of an LLM for decision-making could significantly impact its sustainable strategies, as biases inherent in the training data may translate into analysis and output. 
Consequently, examining how LLMs interpret and respond to sustainability-related concepts is essential, especially as organizations increasingly depend on these technologies to address multifaceted environmental and social issues. 

In this study, we perform an empirical multi-model evaluation to assess the biases of various LLMs regarding the perceived importance of corporate social sustainability practices and stakeholder relationships within supply chains. To determine the practical implications of these biases, we further prompt the LLMs to assume the perspective of an employee representing one of four prominent organizational culture types. Prior research indicates that organizational culture significantly influences the implementation of sustainable strategies, as distinct cultures are associated with varying capacities, motivations, and processes in corporate social responsibility activities \citep{maheshwari2024}. Our analysis examines whether different cultural value orientations amplify or attenuate LLM-generated perspectives on ethics, corporate responsibility, and environmental collaboration with suppliers and customers. The goal is to answer the following research questions:

RQ1: To what extent do LLMs display biases when evaluating the importance of corporate sustainability practices and stakeholder relations in green supply chains?

RQ2: How does prompting LLMs to adopt specific organizational cultures influence their assessments of the importance of corporate sustainability practices and stakeholder relations in green supply chains?

By answering these research questions, we contribute to an interdisciplinary crossroad of research directions, namely, green supply chain management, corporate sustainability, and LLM evaluation/ psychometrics. 
Specifically, this study contributes to theory building and practical recommendations in these ways: First, by uncovering underlying response patterns in different LLMs -- especially in the context of different organizational cultures -- we can make predictions about the extent to which the use of LLMs may have potential unintended or undesirable consequences for different organizations. Second, the results add to the research domain of LLM psychometrics, an important methodological approach for evaluating technological outcomes. 
This work provides significant implications for decision-makers in organizations that increasingly rely on LLMs to support sustainability-related activities, including supplier evaluation, sustainability reporting, and strategic planning.

\section{Theoretical Background}
\subsection{Corporate Social Responsibility and Green Supply Chain}

In recent years, organizations have increasingly prioritized sustainable business practices within their global strategies. This is often captured under the umbrella term corporate social responsibility (CSR; \citet{ahi2013}), referring to organizational initiatives that extend beyond economic and legal requirements to promote social good \citep{McWilliams2001}. CSR encompasses a range of activities, including the enhancement of sustainability and environmental performance \citep{McWilliams2005}. This trend reflects a broader societal paradigm shift toward sustainable development \citep{WCED1987}. 
The United Nations’ 2030 Agenda for Sustainable Development operationalizes sustainability through 17 Sustainable Development Goals (SDGs), which have become critical objectives in organizational strategy and drive stakeholders to systematically incorporate sustainability considerations into decision-making \citep{vannyaroson2024}. 

Central to this work is the United Nations Sustainable Development Goal (SDG) ("Responsible consumption and production"), which aims to promote the sustainable and efficient use of natural resources, to encourage companies to minimize their social and environmental risks, and to better inform consumers about sustainable consumption \citep{BMZ_SDG12}.

Central to this work is the United Nations Sustainable Development Goal (SDG) 12 ("Responsible consumption and production"), with one of its most important levers: supply chain management (SCM). Supply chain management is the coordination and integration of the flow of materials, information, and financial resources across the entire supply chain (from suppliers to customers) to optimize overall system performance \citep{Stroumpoulis.2022}. In the context of sustainability, green supply chain management (GSCM) is discussed as "one of the essential elements for sustainable development and survival" of organizations (\citet{lee2020}, p. 2). 
Defining GSCM presents challenges, as SCM encompasses a wide range of activities and there is considerable variation regarding the components that should be included in both the definition and implementation of GSCM \citep{fahimnia2015, srivastava2007}. 
Broadly conceptualized, GSCM refers to initiatives that aim to incorporate environmental aspects into SCM \citep{lee2020}. This encompasses a wide range of tasks including sourcing (e.g., renewable energies), product design (e.g., use of sustainable material), delivery (e.g., emissions), customer and product use (e.g., customer education), recycling (e.g., return of products) or life cycle assessment \citep{hassini2012, srivastava2007,becker2025forecasting, tokkozhina2025}. GSCM is one aspect of the broader defined sustainable SCM, which not only includes environmental but also societal goals \citep{das2017, ahi2013}. 

The integration of CSR and GSCM practices creates essential foundations for circular economy implementation, particularly in the context of emerging digital technologies. As organizations transition toward circular supply chain models that emphasize closed-loop systems, resource efficiency, and waste minimization \citep{Awan2022}, governance frameworks established through CSR initiatives provide essential stakeholder engagement mechanisms and sustainability standards required for successful circular economy adoption. This alignment becomes increasingly relevant as generative Artificial Intelligence (Gen AI) and other digital transformation technologies offer new capabilities for optimizing circular practices, from predictive analytics for demand forecasting to automated systems for product lifecycle management and waste stream optimization \citep{Li2024generativeAI,Mariani2024}. 
Their responsible implementation within circular economy frameworks requires the kind of comprehensive sustainability governance that CSR-driven GSCM practices establish, ensuring that technological advances serve broader environmental and social objectives rather than purely economic efficiency gains \citep{Chauhan2022}. Furthermore, as Gen AI facilitates greater involvement of multiple circular supply chain actors across the value chain -- including material producers, end-users, and recycling organizations \citep{Mariani2024} -- the stakeholder-oriented approach inherent in CSR becomes critical for managing these complex, interconnected relationships in pursuit of carbon neutrality goals \citep{akhtar2024}.

The proposed advantages of GSCM are multifaceted, ranging from superior environmental and economic performance -- including return on assets as well as core business outcomes\citet{li2019} -- to enhanced reputation or improved fulfillment of customer needs \citep{das2017, tseng2019}. 

\subsection{LLMs in Green Supply Chain Management}
The use of technological innovation is proposed as a great leverage for implementing green supply chain practices successfully at scale \citep{zhou2025}. The parallel advancement of sustainable development and technology innovation is broadly referred to as the "twin pursuit" \citep{felder2025smart} or "twin transition" \citep{Bush.2025}. Technological innovations such as AI, can increase sustainable business practices, e.g., through monitoring, analytical, and decision-making capabilities \citep{Vinuesa2020}. 
In particular, LLMs are increasingly discussed as a catalyst for sustainable business practices \citep{akhtar2024, tokkozhina2025, preuss2024}. LLMs are advanced Gen AI systems (foundation models) capable of natural language (text or speech) processing and generation \citep{10487437}. 
Several use cases of Gen AI in general and LLMs specifically have been discussed for GSCM \citep{dwivedi2023} such as life cycle assessment, the generation of optimal warehouse layout, designing and manufacturing environmentally friendly products, the optimization of logistic decisions, customer support or the creation of personalized content for customers \citep{akhtar2024, tokkozhina2025, preuss2024}.

LLMs can be implemented in GSCM through data analysis for supply chain optimization, predictive modeling for demand forecasting, automated compliance monitoring, and decision support systems that synthesize sustainability criteria into actionable recommendations \citep{Jackson2024,Wamba2023}. Our study focuses on LLM-assisted decision-making, where models serve as intelligent advisors.
For example, a supply chain manager might use an LLM to analyze vast sales data sets for more accurate demand forecasting or for suggesting potential future scenarios based on various decision-making criteria \citep{Aghaei2025}. 

Although there are promising LLM use cases for GSCM, a wide range of implementation challenges are discussed \citep{Aghaei2025}. \citet{preuss2024} summarizes five main risks of LLM use for life cycle assessment -- a subtask of GSCM -- that are 1) transparency of generated content, 2) misinformation and bias, 3) limitations of training data, 4) accountability and responsibility, and 5) sustainability. These risks are not unique to the application of LLMs in SCM, but relate to general challenges of LLM implementation in the business context \citep{tokkozhina2025}. 
In the following, we specifically focus on the risk of LLM biases: 
We emphasize LLM biases because sustainable decision-making directly impacts environmental outcomes and stakeholder welfare across supply chain networks. Biases in LLM recommendations can systematically favor certain suppliers or practices while overlooking marginalized communities or sustainable alternatives, thereby potentially undermining environmental justice principles in GSCM \citep{Budhwar2023}. Since supply chain decisions involve complex economic-environmental-social trade-offs, mitigating bias is essential to ensure that LLM recommendations align with comprehensive sustainability objectives.
In the following, we specifically focus on LLM biases because they can systematically disadvantage marginalized communities and sustainable alternatives while compromising environmental justice principles, thereby impeding the achievement of comprehensive sustainability objectives that balance economic, environmental, and social trade-offs across complex supply chain networks.

\subsection{LLM Biases}
LLM biases have been defined as "the presence of systematic misrepresentations, attribution errors, or factual distortions that result in favoring certain groups or ideas, perpetuating stereotypes, or making incorrect assumptions based on learned patterns" (\citet{Ferrara_2023}, p. 2ff). Demographic biases are widely documented. For example, LLMs tend to associate women with family-related roles \citep{lucy-bamman-2021-gender, Kotek2023}, reproduce negative stereotypes about racial minorities \citep{Khandelwal}, depict older adults as less competent \citep{shin-etal-2024-ask}, or disproportionately link Islam with violence \citep{abid2021muslims}. Cultural and linguistic biases are also prominent, as models trained predominantly on English data often mirror Western-centric values \citep{tao2024culturalbias} and under-represent non-Western perspectives \citep{aksoy2024}. In addition, political biases have been observed, where LLMs favor certain ideologies or political positions \citep{motoki2024more,rutinowski2024self}, or give non-neutral political recommendations \citep{dormuth2025cautionarytaleneutrallyinformative}. 
Beyond these, political and ideological biases have been observed, with studies showing that models can lean toward particular partisan positions depending on prompt framing \citep{elbouanani2025politicalbias}, model family \citep{weber2024gpt4politicallybiasedgpt35}, or structural properties of large-scale training \citep{resnik2025llmbias}.
LLMs exhibit well-documented demographic, cultural-linguistic, and political biases that can systematically favor certain groups, perspectives, and ideologies over others \citep{lucy-bamman-2021-gender, Kotek2023, tao2024culturalbias, aksoy2024, motoki2024more}.

In the context of CSR and GSCM, potential sustainability biases of LLMs are of central concern, i.e., the question of "how LLMs understand and express attitudes towards sustainability principles" (\citet{bush2025emnlp-arxiv}, p. 2). Sustainability biases can be expressed within a specific model, e.g., \citet{Kuehne.2024} discovered a sustainability bias in infrastructure-related queries, with LLMs showing stronger emphasis on social aspects of sustainability while often under-representing economic and environmental components. 
Furthermore, recent empirical work revealed inter-model differences concerning how five state-of-the-art LLMs (Claude, DeepSeek, GPT, LLaMA, and Mistral) conceptualize the compatibility of AI and sustainability \citep{bush2025emnlp-arxiv}.
Such biases are especially critical as users hardly detect biases in AI in decision-making processes which leads to high risks of taking wrong decisions \citep{Kuhl.2025}.
These findings highlight the importance of understanding embedded biases in AI systems when they are deployed for sustainability-related decision-making, e.g., in the context of GSCM.

\subsection{LLM Evaluation and Psychometrics}
Various methodologies have been used to evaluate LLMs: 
LLM benchmarking aims to rank models based on a specific task \citep{ye2025}. 
In practice, benchmarks rely on curated datasets and standardized test suites to elicit model responses and compare them against predefined reference answers or human performance levels. 
Prominent examples include knowledge- and reasoning-oriented tasks such as MMLU \citep{hendrycks2021mmlu}, stereotype-focused benchmarks like WinoBias \citep{zhao2018winobias} and BBQ \citep{parrish2022bbq}, and GEST \citep{pikuliak2023gest}. 
These resources enable reproducible cross-model comparisons by quantifying the extent to which models reproduce stereotypes, demographic imbalances, or cultural assumptions. 
However, benchmark-based evaluations have been criticized for capturing only narrow slices of model behavior and lacking construct validity, as they often abstract away from real-world contexts and quickly become outdated as models evolve \citep{li2023surveyfairness, ye2025}.

Qualitative evaluations focus on the generated LLM output and evaluate it based on e.g., qualitative content analysis or pre-defined metrics (e.g., \citet{Giudici.2023}). 

A quantitative approach to LLM bias evaluation draws on validated psychometric questionnaires. This interdisciplinary research field has been coined "LLM Psychometrics" and is dedicated to not only evaluating but understanding and enhancing LLMs based on the use of psychometric instruments \citep{ye2025}. 
By evaluating how LLMs respond to psychometric questionnaires that have been tested and validated on human samples, researchers have demonstrated, e.g., political biases \citep{motoki2024more,weber2024gpt4politicallybiasedgpt35,dormuth2025cautionarytaleneutrallyinformative}, cultural biases \citep{aksoy2024} and biases towards the compatibility of AI and sustainability \citep{bush2025emnlp-arxiv}. Compared to LLM benchmarking, LLM psychometrics is proposed to provide more generalizable and real-world applicable findings that are less likely to become outdated quickly\citep{ye2025}. 
The approach has been criticized for interpreting LLM outputs as ‘opinions’ or ‘attitudes’, even though these merely reflect statistical patterns \citep{Bender2021}. It is thus important to adopt adequate terminology when using LLM psychometrics \citep{ye2025}. 
Furthermore, it is important to consider if and how these patterns found in LLM responses translate to real-world scenarios \citep{Bender2021,ye2025}. The general assumption is that if LLM responses exhibit distinct patterns that suggest ideological biases, such biases impact real-world applications. 
In real-world applications such as GSCM, a supply chain manager might use an LLM chatbot for decision-making, e.g., concerning the question of which or how much supplier to choose. In reality, there is likely a chat history, e.g., regarding corporate goals or strategic orientations of the organization and LLM responses vary depending on the temporal dynamics \citep{ye2025}. One way to incorporate this dimension into LLM psychometric evaluation is to not only treat LLMs as 'one entity' using default settings, but to evaluate LLMs by using various LLM trait expressions, e.g., by prompting specific personas \citep{ye2025}. This way, different practical settings can be tested for their impact on LLM outputs, thereby analyzing whether previously demonstrated biases translate into these applications. In this study, we consider the context of organizational culture to determine the practical implications of potential sustainability biases.

\subsection{The Role of Organizational Culture}
Organizational culture is a prominent concept in work and organizational psychology. One widely cited definition is that organizational culture constitutes "the shared values and basic assumptions that explain why organizations do what they do and focus on what they focus on" (p. 468, \citet{schneider2017}. 
Organizational culture is an umbrella term for a  holistic concept that tries to capture the "personality" of an organization, consisting of 'shared values and basic assumptions that explain why organizations do what they do and focus on what they focus on' (p. 468, \citet{schneider2017}. Employees interpret and understand the goals, processes work, task components and relationships based on the cultural values that are promoted within an organization \citep{Schneider2013}. 
One prominent model of organizational culture is the Competing Values Framework (CVF) that distinguished four organizational culture types based on two dimensions: flexibility vs. stability and internal vs. external focus \citep{quinn1981}. (1) \textit{Clan cultures} are characterized by flexibility and an internal focus and put emphasize on teamwork, trust and cooperation. (2) \textit{Adhocracy cultures} are characterized by flexibility and an external focus, and put emphasis on innovation, employee empowerment, autonomy and risk-taking. (3) \textit{Market cultures} are characterized by stability and an external focus and put emphasis on achievement, performance, recognition and goal-orientation. Lastly, (4) \textit{hierarchy cultures} are characterized by stability and an internal focus and put emphasis on structure, role clarity, ethics, safety and control \citep{quinn1981, beus2020}. Research has shown that the four dimensions are configurative, do not compete \citep{Hartnell2011}, and can co-exist in organizations. However, some values are likely more prominent than others \citep{QuinnKimberly1984}. Thus, the model is useful as a meta-theory and has received wide support in that regard \citep{beus2020, kluge2002assessment}. 

Organizational culture has been shown to impact corporate sustainability initiatives \citep{maheshwari2024, belay2023}. Although it has been stressed that there is no single 'best' culture for sustainable business practices, their effectiveness depends on how well these strategies align with the respective culture \citep{iqbal2025, lazar2022, dyck2019}. In fact, empirical analyses demonstrate differences between the culture types and CSR practices:  
\textit{adhocracy} and \textit{clan culture} have been positively linked to eco-innovation \citep{shuliang2024} as well as the implementation of GSCM \citep{iddik2024}. In line with that, adhocracy culture has been positively associated with efforts concerning the external integration of key suppliers and customers in SCM \citep{braunscheidel2010}. Although some studies find a positive link between \textit{hierarchy culture} and GSCM, other studies identify it as a barrier in CSR implementation \citep{braunscheidel2010, dawar2023, bortolotti2024}. Lastly, there seems to be no relation between \textit{market culture} and environmental performance \citep{osei2022}. 
These results suggest that depending on the prevailing organizational culture type, approaches towards GSCM will differ. As a real-world application test, we therefore examine whether different cultural value orientations amplify or attenuate LLM-generated perspectives on ethics, corporate responsibility, and environmental collaboration with suppliers and customers.

\subsection{Research Questions}
Existing research has largely focused on general sustainability attitudes rather than specific organizational contexts in which such biases influence consequential decision-making.
As organizations increasingly rely on LLMs for tasks such as supplier evaluation, sustainability reporting, and strategic planning within GSCM frameworks, it becomes essential to understand the systematic patterns in which different models prioritize various aspects of CSR. The convergence of organizational reliance on AI systems for sustainability decision-making and the documented presence of biases in LLMs necessitates a systematic examination of how these models conceptualize and respond to CSR and GSCM principles.
Based on the theoretical background, we identified two primary research questions that are critical for understanding the practical implications of deploying LLMs in sustainability contexts.

\textbf{RQ1: To what extent do LLMs display biases when evaluating the importance of corporate sustainability practices and stakeholder relations in green supply chains?}

\textbf{RQ2: How does prompting LLMs to adopt specific organizational cultures influence their assessments of the importance of corporate sustainability practices and stakeholder relations in green supply chains?}

While RQ1 focuses on identifying the existence and nature of sustainability biases, RQ2 examines whether these biases are amplified, attenuated, or modified when LLMs are prompted to consider specific organizational contexts — a common scenario in practical LLM deployment.

\section{Methodology}
We employ a multi-method psychometric approach to assess how LLMs conceptualize ethics, social responsibility, and environmental collaboration with suppliers and customers. Our evaluation covers five cutting-edge LLMs 
\begin{itemize}
\item\setlength{\itemsep}{0pt} Closed-source models: GPT-4o \citep{openai2024gpt4o} by OpenAI and Claude 3.7 Sonnet \citep{anthropic_claude_sonnet} by Anthropic
\item\setlength{\itemsep}{0pt} Open-source models: LLaMA 3.3 70B-Instruct \citep{meta_llama33_docs} by Meta, Mistral 7B-Instruct \citep{mistralai_large}, and DeepSeek V3 \citep{deepseek2025v3}
\end{itemize}

For GPT-4o and Claude 3.7 Sonnet, responses were generated using their official APIs (OpenAI Python and Anthropic). The open-source models were run under different technical setups: LLaMA 3.3 and Mistral were hosted locally through PyTorch and Hugging Face, while DeepSeek V3 was accessed via its external API. In all cases, we relied on the most recent model releases available during data collection, which took place in August 2025, to ensure that our findings represent their current performance profiles.

\subsection{Psychometric Instruments}
To address RQ1 and RQ2, we employ two psychometric instruments that are well suited for the analysis of ethics, social responsibility, and environmental collaboration:
\begin{enumerate}
    \item Perceived Role of Ethics and Social Responsibility (PRESOR)
\end{enumerate}
The PRESOR scale \citep{Singhapakdi1996} measures perceptions of the importance of ethics and social responsibility for organizational effectiveness. The questionnaire uses a 9-point Likert scale (1 = Strongly Disagree to 9 = Strongly Agree) and captures three dimensions: (1) Social Responsibility and Profitability (4 items, example: "Social responsibility and profitability can be compatible"), (2) Long-term Gains (6 items, e.g., "Business has a social responsibility beyond making a profit."), and (3) Short-term Gains (3 items; e.g., "If the stockholders are unhappy, nothing else matters")

\begin{enumerate}
\item[2.] Green Supply Chain Partnerships (GSCP)
\end{enumerate}
The GSCP scale, originally published by \citet{vachon2006}, measures the degree of environmental collaboration with (a) suppliers and (b) customers. A validated short version \citep{lee2020} captures both dimensions with 4 items each, rated on a 7-point Likert scale (1 = Strongly Disagree to 7 = Strongly Agree). In our study we adapted the items of the short version so that LLM responses could be given in general, i.e., without referring to a specific organization. For example, the original item "we are working together to reduce environmental impact of our activities with our suppliers [customers]" was reformulated as "It is important to work together to reduce environmental impact of our activities with our suppliers [customers]". 

\subsection{Experimental Setup}
The experimental design was structured to guarantee standardization, reproducibility, and statistical rigor. For each LLM and each questionnaire, we employed a uniform system prompt instructing the models to respond strictly in the designated Likert scale format. For instance, in the PRESOR questionnaire the prompt stated: "For each statement, indicate how well it describes you or your opinions. Select one of the following options: Strongly Disagree, Disagree, Somewhat Disagree, Slightly Disagree, Neutral, Slightly Agree, Somewhat Agree, Agree, Strongly Agree."

To prevent deviations from the scale, we imposed additional constraints within the prompts:  
\begin{enumerate}
\item\setlength{\itemsep}{0pt}Provide no reasoning or explanation.  
\item\setlength{\itemsep}{0pt} Respond only with the specified options. \item\setlength{\itemsep}{0pt} Do not apologize or add disclaimers.  
\item \setlength{\itemsep}{0pt}Avoid words such as "cannot", "unable", "instead", "as", "however", "unfortunately" or "important". 
\item\setlength{\itemsep}{0pt} Refrain from producing negative sentences about the prompt itself. 
\end{enumerate}  
These rules were adapted to the structure of each questionnaire to ensure consistency across instruments. 
To capture variability and enable robust statistical analysis, each questionnaire was administered $100$ times per model, yielding $500$ complete response sets per instrument. This setup allowed examination of both central tendencies and response variance across models. 
Data collection scripts and datasets are available in the project’s GitHub repository \footnote{\url{https://github.com/anonim705/LLM_PRESOR_GSCP}}. 

\paragraph{Organizational Culture} 
In a subsequent step, we prompted the LLMs to adopt four different roles pertaining to four prominent organizational cultures. We based the role prompts on the Competing Values Framework \citep{quinn1981} and LLMs were prompted "You are an employee of an organization with a Clan [Adhocracy, Market, Hierarchy] culture". Subsequently, we described each culture in each respective prompt, drawing on \citet{quinn1981} as well as lexical descriptions of the four culture types by \citet{voss2022}. All prompts can be found in \ref{sec:appendixA}.

\subsection{Analyses}
In a first step, we examined the reliability (Cronbach's alpha) of the scales, and their factor structure using confirmatory factor analysis. We conducted these analyses both for the scales in general and separately for the datasets generated based on the four organizational culture role prompts.

To determine overall differences among the five models on the psychometric measures (general assessment) as well as the specific pairs that differ, we employed nonparametric multiple contrast testing procedure (MCTPs) for one-way designs provided by the R package \texttt{nparcomp} \citep{konietschke2015}. All pairwise model comparisons per questionnaire were examined using Tukey-type contrasts, with critical values obtained from a multivariate t-distribution using a Satterthwaite approximation to control the family-wise error rate, thereby adjusting for multiplicity.

For the two-factorial analysis involving both organizational culture and LLM model, we inferred main and interaction effects using rank-based nonparametric ANOVA-type (ATS) tests \citep{brunner2017rank} implemented in the R package \texttt{rankFD} \citep{rankFD}. 

Both rank-based approaches do not require distributional assumptions and quantify group differences in terms of nonparametric relative effects. Given the exploratory nature of our study, we did not apply multiplicity adjustments.

\section{Results}
\subsection{Reliability, Factor Structure and Comparison to Human Data}

As the questionnaires were originally developed for human participants, we first examined the factor structure and reliability of the scales for LLM responses. 

\paragraph{PRESOR scale} Confirmatory factor analysis showed a poor model fit for the three-factor structure of the PRESOR questionnaire ($\tilde{\chi}^2 = 435.22$, $p < .001$, $CFI = .81$, $TLI = .76$, $RMSEA = .11$, $SRMR = .09$), i.e., recommended thresholds were not met. Standardized factor loadings were highly variable, ranging from $\lambda = .09$ to $\lambda = .72$ for Factor 1 (Profitability) and from $\lambda = .17$ to $\lambda = .72$ for Factor 2 (Long-term gain). Factor loadings for Factor 3 (Short-term gain) could not be estimated, likely due to convergence issues resulting from negative inter-item correlations. 

The internal consistency was medium to low for all three subscales in general; when analyzing the organizational culture role prompted subdata sets, Cronbach's alpha values were medium to very low (see Table~\ref{tab:cfa_presor}).

Although there is previous work with human participants that could not reproduce the three-factor structure \citep{etheredge1999perceived}, other studies with human participants successfully replicated and validated the factor structure in different national cultures \citep{shafer2007values, lee2011attitudes}. In the original validation study of the PRESOR scale, Cronbach's alpha values were within an acceptable range for early stage research with this scale (factor 1:  $\alpha = .71$, factor 2:  $\alpha = .57$, factor 3:  $\alpha = .64$; \citep{Singhapakdi1996} and Cronbachs alpha values above .70 have been reported in other studies using the scale (e.g., \citep{zhu2014corporate}).

Thus, it is likely that LLMs were not able to mimic human behavior on the PRESOR scale. Due to the low internal consistencies and large discrepancies between the datasets for each organizational culture type, subsequent analysis with composite subscale scores were not feasible. To answer the research questions concerning the perceived role of ethics and responsibility, we thus proceeded with analyzing one single item per subscale for the PRESOR questionnaire. We chose the items with the highest factor loadings: scale 1 ("Social responsibility and profitability can be compatible", in the following "Profitability"), scale 2 ("Being ethical and socially responsible is the most important thing a firm can do", in the following "long-term goal") and scale 3 ("If the stockholders are unhappy, nothing else matters", in the following "short-term goal").

\paragraph{GSCP scale} For the GSCP questionnaire, confirmatory factor analysis showed a medium to good fit ($\tilde{\chi}^2 = 68.82$, $p < .001$, $CFI = .98$, $TLI = .96$, $RMSEA = .07$, $SRMR = .03$). All items loaded strongly on their respective factors, with standardized factor loadings ranging from $\lambda = .68$ to $\lambda = .76$ for Factor 1 (Relationship with Suppliers) and form $\lambda = .66$ to $\lambda = .80$ for Factor 2 (Relationship with Customers). Internal consistency was satisfactory for most datasets, with the exception of the data generated under the \textit{clan} (both Factors) and \textit{hierarchy} role (Factor 2) prompt assignment, which should therefore be interpreted with caution (see Table~\ref{tab:cfa_presor}). Overall, the scales showed internal consistency comparable to previous findings with human participants. For example, \citet{lee2020} reported Cronbach's alpha values of $.70$ for the first factor and $.64$ for the second factor. 

\begin{table}[htbp]
\centering
\scriptsize
\caption{Internal Consistency (Cronbach’s alpha) for the PRESOR and GSCP Scale.}
\renewcommand{\arraystretch}{1.2}
\begin{tabular}{|l|c|c|c|}
\hline
 & \textbf{Factor 1} & \textbf{Factor 2} & \textbf{Factor 3} \\
\hline
\multicolumn{4}{|l|}{\textbf{PRESOR: Cronbach’s Alpha}} \\
\hline
Overall & .77 & .66 & .25 \\
Clan & .68 & .53 & NA \\
Adhocracy & .46 & .19 & NA \\
Market & .23 & -.08 & .35 \\
Hierarchy & .03 & -.56 & .003 \\
\hline
\multicolumn{4}{|l|}{\textbf{GSCP: Cronbach’s Alpha}} \\
\hline
 & Factor 1 & Factor 2 & -- \\
\hline
Overall & .82 & .82 & -- \\
Clan & .35 & .21 & -- \\
Adhocracy & .86 & .72 & -- \\
Market & .49 & .83 & -- \\
Hierarchy & .82 & .26 & -- \\
\hline
\end{tabular}
\label{tab:cfa_presor}
\end{table}

\subsection{Perceived Role of Ethics and Social Responsibility}
\subsubsection{General assessment} With mean ratings above 7 (scale 1-9), all models mirrored high expectations regarding the compatibility of social responsibility and profitability (see Table~\ref{tab:presor_results}). 
Models also uniformly provided high ratings regarding long-term goals, i.e., the importance of being ethical and socially responsible for an organization. Notably, LLaMA provided perfect scores regarding this long-term goal; the least positive ratings were provided by Claude. 
Model assessments diverged regarding the question of short-term goals ("if the stockholders are unhappy, nothing else matters"). Claude, GPT, and Mistral largely disagreed with this statement, whereas LLaMa and DeepSeek tended to agree (fsee Table~\ref{tab:presor_results}.

\begin{table*}[htbp]
\centering
\scriptsize
\caption{Mean and standard deviations (SD) of LLMs for the three PRESOR items.}
\label{tab:presor_results}
\renewcommand{\arraystretch}{1.2}
\begin{tabular}{lcccccc}
\hline
\textbf{LLMs} & \multicolumn{2}{c}{\textbf{Profitability}} & \multicolumn{2}{c}{\textbf{Long-term Gain}} & \multicolumn{2}{c}{\textbf{Short-term Gain}} \\
\cline{2-7}
              & Mean & SD & Mean & SD & Mean & SD \\
\hline
GPT     & 8.02 & 0.32 & 7.33 & 0.78 & 2.69 & 0.53 \\
Claude  & 7.86 & 0.43 & 6.82 & 0.58 & 1.81 & 0.39 \\
DeepSeek& 8.00 & 0.00 & 8.69 & 0.46 & 6.96 & 0.40 \\
LLaMA   & 8.83 & 0.38 & 9.00 & 0.00 & 7.43 & 0.76 \\
Mistral & 7.12 & 0.83 & 7.03 & 1.16 & 2.86 & 1.50 \\
\hline
\end{tabular}
\end{table*}


Overall global tests showed significant results for the profitability item ($p < .0001$), long-term gain ($p < .0001$), and short-term gain ($p < .0001$).
Regarding the compatibility of social responsibility and profitability, all pairwise model comparisons were significant at the $5\%$ level, except for the comparison of DeepSeek and GPT ($p = .93$). 
For the long-term gain item, Mistral and GPT ($p = .46$) and Claude and Mistral ($p = .14$) did not differ significantly; all other pairwise comparisons were statistically significant at the $5\%$ level.
For the short-term goal item, all pairwise comparisons were significant ($p < .001$) except for Mistral vs. GPT ($p = .99$).

\subsubsection{The role of organizational culture} 
\begin{figure*}
    \centering
    \includegraphics[width=\textwidth]{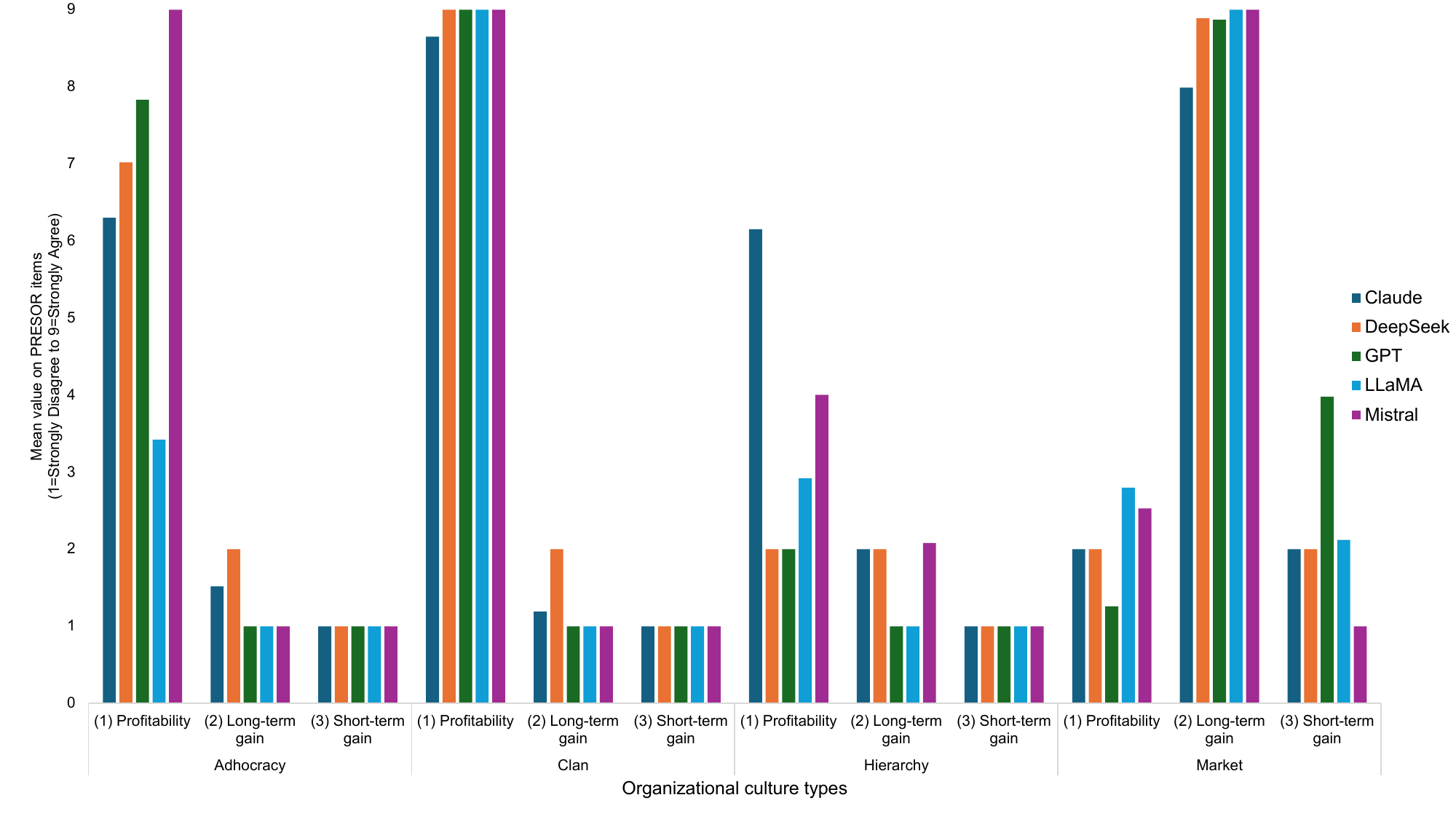}
    \caption{Mean model ratings of PRESOR items for each organizational culture type.}
    \label{fig:Figure_PRESOR_cultures}
\end{figure*}

\autoref{fig:Figure_PRESOR_cultures} gives an overview of how the models evaluated the three aspects of ethical and social responsibility in the role of the four different organizational culture types. 
It is apparent that role assignments lead to differences within and between models and cultures, as ratings showed large differences (\autoref{fig:Figure_PRESOR_cultures}) -- especially compared to the fairly uniform ratings without role assignments.

Rank-based two-way ANOVA tests showed significant main effects of culture (profitability: ATS = 12416.83, p < .0001, long-term gains: ATS = 6794.05, p < .0001, short-term gains: ATS = 3450.39, p < .0001) and model (profitability: ATS = 307.15, p < .0001, long-term gains: ATS = 866.88, p < .0001, short-term gains: ATS = 357.69, p < .0001) for all three subscales. Thus, there were significant differences in LLM outputs between at least two culture types (main effect of culture) and at least two LLMs (main effect of model). 
The interaction effect between culture and model was also significant for all three items (profitability: ATS = 328.49, p < .0001, long-term gains: ATS = 200.19, p < .0001, short-term gains: ATS = 357.69, p < .0001).

\subsection{Stakeholder Relations in Green Supply Chains}
\subsubsection{General assessment} Descriptively, the overall assessment of all five models again demonstrated rather high ratings regarding the importance of environmental cooperation with suppliers and customers (for descriptive information see Table~\ref{tab:gscp_results}. Lowest ratings were given by DeepSeek and highest ratings by Claude. Notably, Claude provided nearly-perfect ratings (maximum of the scale: 7). The overall global test showed a significant result for the collaboration with supplier ($p < .001$) as well as collaboration with customers scale ($p < .001$). All pairwise comparisons proved to be significant for both scales, meaning that the nuances between model ratings were statistically significant ($p < .001$). This result is likely due to the very low variances in ratings within each LLM.

\begin{table}[htbp]
\centering
\scriptsize
\caption{Mean and standard deviations (SD) of LLMs for GSCP subscales.}
\label{tab:gscp_results}
\renewcommand{\arraystretch}{1.2}
\begin{threeparttable}
\begin{tabular}{lcccc}
\hline
\textbf{LLMs} & \multicolumn{2}{c}{\textbf{RWS}} & \multicolumn{2}{c}{\textbf{RWC}} \\
\cline{2-5}
                 & Mean & SD & Mean & SD \\
\hline
GPT  & 6.73 & 0.21 & 6.44 & 0.25 \\
Claude   & 6.98 & 0.07 & 6.97 & 0.08 \\
DeepSeek & 5.68 & 0.15 & 5.44 & 0.22 \\
LLaMA    & 5.85 & 0.36 & 5.77 & 0.34 \\
Mistral  & 6.01 & 0.23 & 6.00 & 0.22 \\
\hline
\end{tabular}
\begin{tablenotes}
\small
\item Note. RWS = Relationship with Suppliers ; RWC = Relationship with Customers.
\end{tablenotes}
\end{threeparttable}
\end{table}

\subsubsection{The role of organizational culture}
\begin{figure*}
    \centering
    \includegraphics[width=\textwidth]{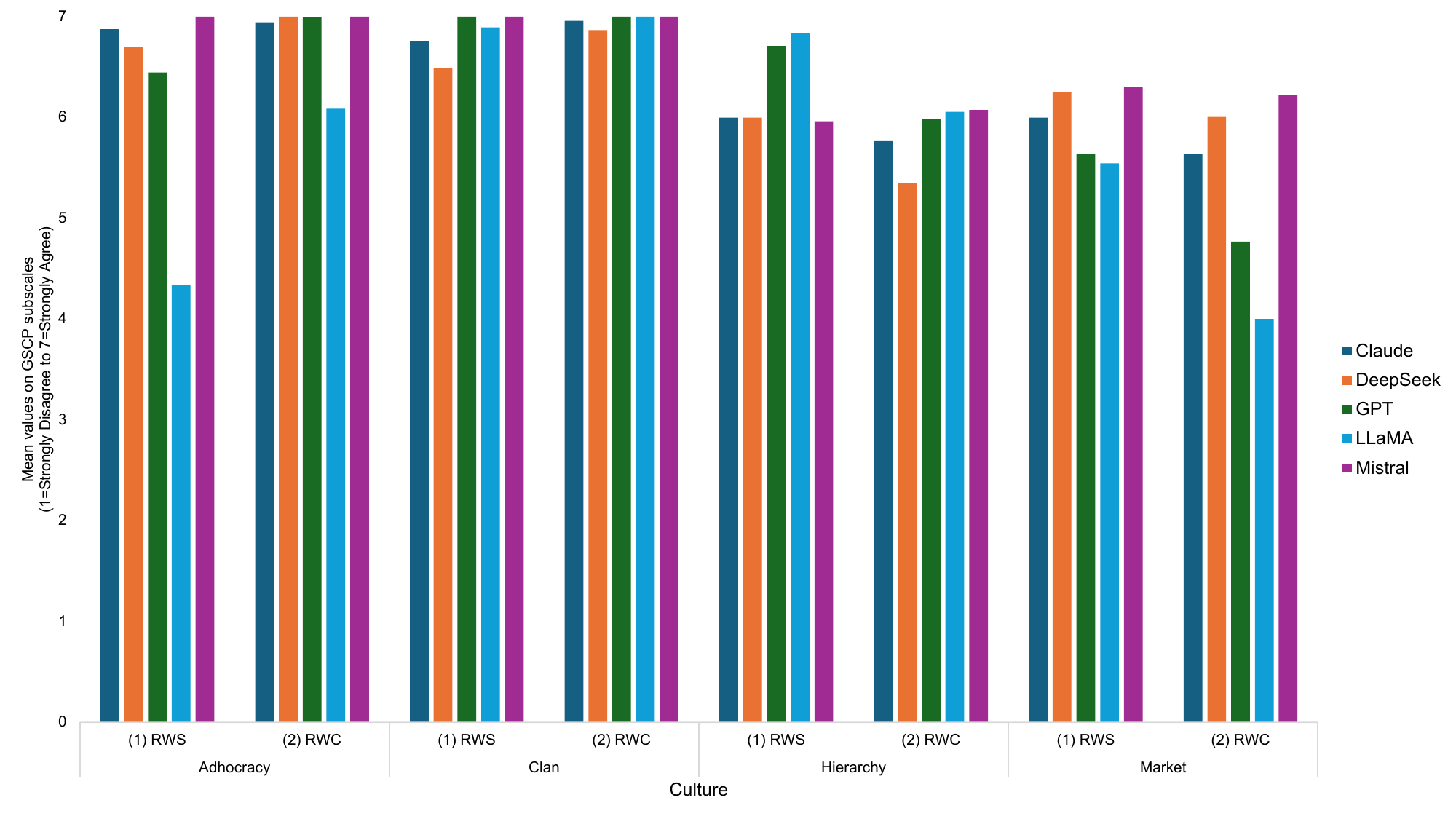}
    \caption{Mean model ratings of GSCP subscales for each organizational culture type.}
    \label{fig:Figure_GSCP_Cultures}
\end{figure*}

\autoref{fig:Figure_GSCP_Cultures} gives a first overview of the model ratings for the two GSCP subscales for each organizational culture type. Descriptively, model ratings became more nuanced and diverging within and between the four culture types, compared to the overall assessment by the models. 
Model ratings became more nuanced and diverging within and between the four culture types for the two GSCP subscales, compared to the overall assessment by the models (\autoref{fig:Figure_GSCP_Cultures}). 

The rank-based two-way ANOVA test showed a significant main effect of culture for the relationship with suppliers ($ATS = 3485.26$, $p < .0001$) as well as the relationship with customers scale ($ATS = 7960.54$, $p < .0001$). The main effect of model was also significant for both scales (suppliers: $ATS = 277.23$, $p < .0001$; customers: $ATS = 401.20$, $p < .0001$). Lastly, the interaction effect was also significant for the relationship with suppliers ($ATS = 1085.31$, $p < .0001$) and customers scale ($ATS = 356.05$, $p < .0001$).

\section{Discussion}

In this study, we aimed to identify if five state-of-the-art LLMs pertain underlying training data biases regarding the importance and role of sustainable business practices and GSCM practices. 
We further examined how answer patterns differed if the LLMs were to assume the role of an employee in one of four prominent organizational culture types.
The findings reveal significant systematic differences between models and demonstrate that organizational culture prompts substantially modify LLM responses.
This holds important implications for LLM-assisted decision-making in sustainability contexts.

\subsection{LLM Biases in Sustainability Assessment (RQ1)}
\subsubsection{Perceived role of ethics and social responsibility}

The analysis of PRESOR responses revealed distinct patterns of biases across the five LLMs examined. All models demonstrated strong agreement with the compatibility of social responsibility and profitability, suggesting an embedded optimistic perspective regarding sustainable business practices. However, differences emerged in the strength of this conviction, with LLaMA providing the most positive assessments and Mistral showing more moderate enthusiasm.

More pronounced biases appeared in responses to short-term versus long-term business orientations. Although all models endorsed the importance of being ethical and socially responsible as a long-term goal, they diverged sharply on shareholder primacy. Claude, GPT, and Mistral strongly rejected the statement ``if stockholders are unhappy, nothing else matters,'' whereas LLaMA and DeepSeek showed substantial agreement with it. These findings suggest fundamental differences in how models conceptualize the relationship between financial performance and stakeholder interests, potentially reflecting different training data sources or fine-tuning approaches \citep{resnik2025llmbias, Ferrara_2023}.

The strong rejection of shareholder primacy by some models (particularly Claude) may indicate an embedded bias toward stakeholder capitalism, while the acceptance by others (particularly LLaMA and DeepSeek) suggests a more traditional shareholder-focused perspective. These differences align with previous findings of political and ideological biases in LLMs \citep{weber2024gpt4politicallybiasedgpt35,rutinowski2024self,elbouanani2025politicalbias}, extending these biases to corporate governance philosophies.

\subsubsection{Green supply chain partnerships}
For environmental collaboration with suppliers and customers, all models provided high ratings, indicating a general bias toward supporting green supply chain initiatives. However, despite these high ratings, pairwise comparisons showed significant differences at the $5\%$ level, thus revealing subtle but consistent differences between models. Claude's near-perfect scores suggest an extremely strong pro-environmental bias, while DeepSeek's comparatively lower (though still positive) ratings indicate a more measured approach to environmental collaboration. These findings align with previous research showing that LLMs can exhibit different sustainability-related biases \citep{bush2025emnlp-arxiv} and complement the findings of documented demographic and cultural biases in other domains \citep{lucy-bamman-2021-gender, Khandelwal, tao2024culturalbias}.

Yet, inter-model differences should not be overstated for this general assessment, as they might primarily result from low variance and rather large sample sizes. Thus, we rather emphasize the consistent pattern of high ratings across models: this suggests that environmental collaboration is generally viewed positively in the training data of contemporary LLMs, potentially reflecting the increasing prominence of sustainability discourse in business literature and corporate communications. This pattern is consistent with the broader finding that LLMs tend to reproduce values and perspectives present in their training data \citep{gallegos2024biasfairnesslargelanguage, aksoy2024}, in this case reflecting the growing emphasis on environmental responsibility in organizational contexts.

\subsection{The Moderating Role of Organizational Culture (RQ2)}
\subsubsection{Cultural context as a bias amplifier}
Introducing organizational culture substantially altered LLM responses, revealing the adaptable nature of these apparent biases. The significant main effects of culture and model, and their interaction across all measures demonstrate that LLM responses to sustainability questions are highly context-dependent. These findings extend previous research on the role of prompting strategies in shaping LLM outputs \citep{kamruzzaman2024prompting, brucks2025prompt} to the domain of organizational sustainability assessment.

For the compatibility of social responsibility and profitability, \textit{clan culture} generated the highest agreement scores between models, which contrasted sharply with \textit{market culture} that produced the lowest scores. This pattern suggests that when prompted to adopt a collaborative and family-like organizational perspective, LLMs emphasize harmony between social and financial objectives. In contrast, when adopting a competitive, results-driven market orientation, models become more skeptical of this compatibility, potentially viewing sustainability initiatives as constraints on performance maximization. A similar pattern emerged for the GSCP dimensions: When prompted to assume the role of an employee in a family-oriented, collaborative \textit{clan culture}, the importance of establishing relations with suppliers and customers for GSCM was valued more highly compared to when assuming a performance-oriented \textit{market culture}. These findings align with empirical research showing that \textit{clan} and \textit{adhocracy cultures} are positively linked to eco-innovation and GSCM implementation \citep{shuliang2024, iddik2024}, while \textit{market cultures} may create tensions with environmental objectives.

The long-term versus short-term orientation items revealed particularly striking cultural effects. \textit{market culture} prompted substantially higher agreement with being ethical and socially responsible as organizational priorities, while other cultures showed much lower support. This counterintuitive finding suggests that when adopting a performance-focused perspective, LLMs may interpret ethical behavior as instrumental to competitive success rather than intrinsically valuable. This result contrasts with research suggesting that there is no relation between \textit{market culture} and environmental performance \citep{osei2022}, indicating that LLMs may embed different assumptions about the strategic value of sustainability than observed in empirical organizational studies. The significant interaction effects reveal that the same LLM can provide markedly different sustainability assessments depending on organizational context. For example, response patterns by LLaMA suggested compatibility of profitability and social responsibility for \textit{clan cultures}, but not for \textit{adhocracy}, \textit{hierarchy} or \textit{market cultures}. In contrast, Mistrals ratings suggested high compatibility of profitability and sustainability for \textit{adhocracy} and \textit{clan cultures}, but not for \textit{hierarchy} and \textit{market cultures}. For the relationship with suppliers and customers scales, the most pronounced intra-model differences emerged for LLaMA: whilst model responses within \textit{clan} and \textit{hierarchy cultures} produced high ratings, ratings for \textit{adhocracy} and \textit{market cultures} were substantially lower. 

Taken together, the results have profound implications for forecasting unintended or undesired consequences of LLM-assisted sustainable decision-making \citep{healy2012unanticipated}. As previous research has delineated, unintended consequences of the digital transformation of the sustainable circular economy might be neutral, positive, or negative \citep{chung2025sustainable}. The results of this work add to this, as they demonstrate how LLMs differ in their embedded perspectives on corporate sustainability practices and stakeholder relations in green supply chains. Especially important in this regard are the documented interaction effects between culture and model. Documenting such biases is the first step to enable a realistic assessment of how LLM-decisions might shape organizational strategy in the long run -- in a neutral, positive, or negative way.

\subsection{Methodological Insights: LLM Psychometrics in Sustainability Research}
The contrasting performance of PRESOR (poor reliability) versus GSCP (acceptable reliability) highlights that human-validated instruments may not maintain psychometric properties when applied to LLMs, requiring careful validation of psychological assessment tools for LLM applications \citep{ye2025, li2023surveyfairness}. Yet, it must be noted that there is also previous work with human participants that did not replicate the factor structure of the PRESOR scale \citep{etheredge1999perceived}, hinting at the possibility that psychometric problems in this study were due to conceptual issues of the scale rather than fundamental differences between human and AI-generated answers. However, as the scale has also been replicated multiple times before, it is likely that that LLMs may process sustainability concepts differently from humans, with implications for interpreting and generalizing from AI psychometric assessments.

The successful use of organizational culture prompts demonstrates the value of context-manipulation approaches in LLM research. The significant model-culture interactions indicate that understanding AI bias requires examining how model characteristics interact with specific application contexts \citep{ye2025}. This methodology addresses critiques about real-world applicability by testing how biases manifest under different organizational scenarios, providing a bridge between laboratory assessment and practical deployment contexts in sustainability decision-making.

\subsubsection{Practical implications}
As the ``twin transition'' \citep{Bush.2025} toward sustainability and digitization accelerates, understanding and mitigating AI biases in environmental decision-making becomes increasingly critical. 
The documented biases have profound implications for organizations that increasingly rely on LLMs for sustainability-related tasks such as supplier evaluation, sustainability reporting, and strategic planning \citep{zhou2025, dwivedi2023, akhtar2024, tokkozhina2025}. 

The substantial effects of organizational culture prompts demonstrate that LLM sustainability perspectives are highly context-dependent \citep{resnik2025llmbias, kamruzzaman2024prompting}. This enables a more nuanced assessment but also raises concerns about consistency and potential manipulation. For example, a company with a strong \textit{market culture} using Claude might receive different sustainability recommendations as if they used LLaMA. Another example evident from the analysis is that organizations that use different LLMs for strategic advice may receive fundamentally different recommendations regarding stakeholder prioritization. The significant interaction effects with organizational culture on top mean that organizations may inadvertently receive recommendations aligned with existing cultural biases rather than objective assessments \citep{Ferrara_2023, gallegos2024biasfairnesslargelanguage}. 
The findings demonstrate that model selection itself becomes a strategic decision with sustainability implications. Thus, organizations must carefully consider the embedded biases of different AI systems and the alignment of those perspective with their respective business strategies when deploying them for environmental decision-making. Decision-makers in organizations and policymakers must recognize that AI systems are not neutral tools but rather embody specific values and assumptions that can systematically influence sustainability outcomes. 
This calls for greater attention to bias auditing, model selection criteria, and governance frameworks that ensure AI-assisted sustainability decisions align with broader environmental and social objectives.

\subsection{Limitations}




This study has several limitations that suggest directions for future research. 
The focus on Western organizational culture frameworks and English-language prompts may limit the generalizability of findings across linguistic and cultural contexts. 
Future research should examine how cultural prompts reflecting different national cultures influence LLM sustainability assessments.

The selection of models, while comprehensive, reflects a particular moment in AI development. The rapid evolution of LLM capabilities and training approaches means that the findings may have limited temporal generalizability. Longitudinal studies tracking how sustainability biases evolve as models are updated and retrained would provide important insights into the stability of these patterns over time.

The use of single-item measures for PRESOR due to reliability issues limits the depth of analysis for ethical and social responsibility constructs. 
Future research should develop and validate LLM-specific instruments to assess sustainability-related attitudes and biases.

Furthermore, this study examined LLM responses in isolation from real-world decision-making contexts. Research examining how these biases influence actual organizational processes and human decision-making would provide valuable insights into the practical significance of the documented patterns. Future research could build on this work and take a qualitative evaluation approach to analyze and compare response patterns across more complex and realistic tasks.

\section{Conclusion}

This study systematically examined how LLMs conceptualize and respond to CSR and GSCM principles, revealing significant biases that have important implications for organizational sustainability decision-making. 
Through a comprehensive multi-model evaluation using validated psychometric instruments, we demonstrated that contemporary LLMs exhibit systematic patterns of bias when assessing sustainability practices and stakeholder relationships, and that these biases are substantially moderated by organizational cultural contexts.

Our analysis of five state-of-the-art LLMs revealed critical findings that advance understanding of AI bias in sustainability contexts. 
All models demonstrated a general pro-sustainability bias, consistently providing high ratings for environmental collaboration and social responsibility initiatives. However, significant inter-model differences emerged in fundamental business philosophy orientations, particularly regarding the tension between shareholder primacy and stakeholder capitalism. 

Current LLMs embed particular perspectives on sustainability-business relationships that may not be universally applicable. 
While the observed pro-sustainability bias could benefit environmental outcomes, it raises concerns about balanced decision-making in contexts with genuine economic-environmental trade-offs \citep{preuss2024, Kuhl.2025}. 

The systematic patterns revealed in this study suggest that effective LLM-assisted sustainability management requires not only technical sophistication, but also careful attention to the values and assumptions that LLMs bring to environmental decision-making. 
As we continue to integrate LLMs into sustainability governance frameworks, maintaining awareness of these biases while working to develop more balanced, representative LLMs becomes essential for achieving genuine progress toward environmental and social objectives.

\vspace{1em}
\noindent\textbf{Declaration of Generative AI and AI-assisted Technologies in the Writing Process}\\
Generative AI tools were used to suggest non-substantive R code edits. The authors reviewed and verified all R code and outputs. No data were shared with the tool. Generative AI tools were also used for translation and language editing to enhance readability. The authors subsequently reviewed and revised the output as needed and take full responsibility for the final content.

\bibliographystyle{elsarticle-harv} 
\bibliography{references}

\appendix

\section{Organizational culture role prompts}
\label{sec:appendixA}
\begin{itemize}
 \item \textbf{Clan:} You are an employee of an organization with a Clan culture. This type of organization has an internal focus and values flexibility. It is structured like a family, emphasizing cooperation, trust, and employee commitment. Your responses should reflect a culture that values cooperation, involvement, teamwork, trust, and care for employees.
    
 \item \textbf{Adhocracy:} You are an employee of an organization with an Adhocracy culture. This type of organization has an external focus and values flexibility. It is a dynamic, entrepreneurial, and innovative environment with an emphasis on risk-taking and experimentation. Your responses should reflect a culture that values innovation, empowerment, autonomy, risk-taking, and creativity.
    
  \item \textbf{Market:} You are an employee of an organization with a Market culture. This type of organization has an external focus and values stability. It is a results-driven, competitive atmosphere with a focus on goal achievement, productivity, and market share. Your responses should reflect a culture that values achievement, performance, work pressure, recognition, and goal orientation.
    
   \item \textbf{Hierarchy:} You are an employee of an organization with a Hierarchy culture. This type of organization has an internal focus and values stability. It is a formalized, structured, and rule-driven environment with an emphasis on efficiency, consistency, and predictability. Your responses should reflect a culture that values structure, role clarity, ethical aspects, safety, and control.
\end{itemize}
\end{document}